\documentclass[12pt, draftclsnofoot, onecolumn]{IEEEtran}

\ifCLASSINFOpdf
\else
   \usepackage[dvips]{graphicx}
\fi
\usepackage{url}
\usepackage{cite}
\usepackage{amsmath,amssymb,amsfonts}
\usepackage{algorithmic}
\usepackage{graphicx}
\usepackage{textcomp}
\usepackage{xcolor}
\usepackage{numprint}
\usepackage[squaren,Gray]{SIunits}
\usepackage{subcaption}
\usepackage{mathtools}
\usepackage{mathrsfs}
\usepackage{stmaryrd} 
\usepackage{caption}
\usepackage{subcaption}
\usepackage{diagbox}
\usepackage{comment}

\usepackage{amsmath}
\usepackage{amssymb}
\usepackage{multirow}
\usepackage{makecell}

\usepackage{placeins}
\usepackage{booktabs}

\usepackage{tikz}
\usepackage{xcolor}
\usepackage[linesnumbered,ruled,vlined]{algorithm2e}

\SetCommentSty{mycommfont}
\usepackage{stackengine}

\newcommand{\eg}{\textit{e}.\textit{g}.}
\newcommand{\ie}{\textit{i}.\textit{e}.}

\begin{document}

\bstctlcite{IEEEexample:BSTcontrol}

\title{RCNet: $\Delta\Sigma$ IADCs as Recurrent AutoEncoders}

\author{Arnaud~Verdant, William~Guicquero and Jérôme~Chossat
\thanks{Arnaud~Verdant, William~Guicquero are with CEA-LETI, Univ. Grenoble Alpes, France (arnaud.verdant@cea.fr). Jérôme~Chossat is with STMicroelectronics, Grenoble, France. This work is part of the IPCEI Microelectronics and Connectivity and was supported by the French Public Authorities within the frame of France 2030.}}


\maketitle

\begin{abstract}
This paper proposes a deep learning model (RCNet) for Delta-Sigma ($\Delta\Sigma$) ADCs. Recurrent Neural Networks (RNNs) allow to describe both modulators and filters. This analogy is applied to Incremental ADCs (IADC). High-end optimizers combined with full-custom losses are used to define additional hardware design constraints: quantized weights, signal saturation, temporal noise injection, devices area. Focusing on DC conversion, our early results demonstrate that $SNR$ defined as an Effective Number Of Bits (ENOB) can be optimized under a certain hardware mapping complexity. The proposed RCNet succeeded to provide design tradeoffs in terms of $SNR$ ($>$13bit) versus area constraints ($<$14pF total capacitor) at a given $OSR$ (80 samples). Interestingly, it appears that the best RCNet architectures do not necessarily rely on high-order modulators, leveraging additional topology exploration degrees of freedom.

\end{abstract}

\begin{IEEEkeywords}
Incremental Analog to Digital Converter, Delta-Sigma, Recurrent Neural Networks, Compter-Aided Design.
\end{IEEEkeywords}

\section{Introduction}
\label{sec:Intro}

\IEEEPARstart{T}{he} emergence of versatile deep learning frameworks is leading to the rise of data-driven optimization techniques. The human expertise therefore consists in stating the problem, defining the search space and setting input-output mappings. It also focuses on the computational graph, loss functions and regularization terms. This article deals with $\Delta\Sigma$ IADC architectures, relying on this hardware-aware design approach.

\subsection{RCNet: the proposed RAE-IADC analogy}
\label{subsec:DLADC}
\begin{figure}[!h]
\centering
\includegraphics[trim={0mm 90mm 0mm 0mm},clip,scale=0.45]{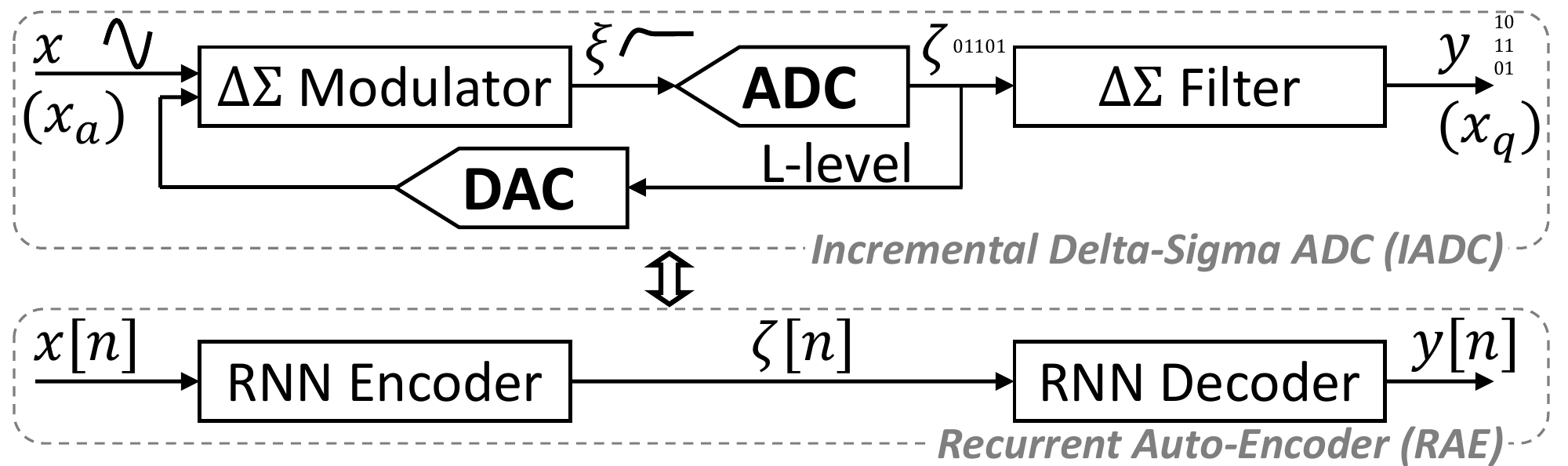}
\caption{RCNet: Analogy between Incremental $\Delta\Sigma$ ADC (IADC) and Recurrent Auto-Encoder (RAE).}
\label{RAE_analogy}
\end{figure}

The presented RCNet methodology relies on the analogy between an IADC based on $\Delta \Sigma$ modulation \cite{markus_incremental_2006, tan_incremental_2020, markus_theory_2004, de_la_rosa_sigma-delta_2011} and a Recurrent AutoEncoder (RAE) (see Fig. \ref{RAE_analogy}). RAE are efficient for a wide range of processing tasks, such as outlier detection in time series \cite{kieu_outlier_2019} \cite{xie_anomaly_2023}, denoising \cite{weninger_deep_2014}, classification \cite{hou_lstm-based_2020} or even compression \cite{yang_learning_2021}. This methodology requires defining the hardware exploration space of the RAE model. RAE training for specific conversion task involves to translate the hardware performance metrics (\eg, quantization error), the internal signal constraints (\eg, dynamic range excursion), and the design robustness specifications (\eg, against noise and components mismatches) into the deep learning framework as fidelity and regularization losses \cite{wang_comprehensive_2022}, activation functions \cite{dubey_activation_2022}, weights quantization functions \cite{thiruvathukal_survey_2022} and data-augmentation layers \cite{mumuni_data_2022}.

\subsection{Prior work on deep learning for ADC}
\label{subsec:DLADC}
Deep learning constitutes a promising avenue to improve performance and robustness for advanced mixed-signal circuit design. Several approaches have already been proposed for Analog-to-Digital Converters (ADC). Indeed, AI-assisted techniques can ease to mitigate hardware non-idealities in \cite{Xu2018AnalogtodigitalCR, samiee_deep_2021, bansal_neural-network_2021}, or to optimize system-level parameters in \cite{de_la_rosa_ai-assisted_2022, fayazi_applications_2021, nam_machine-learning_2021}. Neural Networks (NN) inspired topologies were also proposed \cite{tank_1986}, where an analog network is optimized to approximate any quantization function that provides a binary encoding of the analog input. Additional works such as \cite{cao_neuadc_2020}, \cite{james_neural_2018} have also investigated the approximation of any quantization function while taking into account hardware nonidealities into the training stage. Neuromorphic data converters using memristor arrays even enable mismatch self-calibration \cite{danial_pipelined_2020, danial_delta-sigma_2019}.

Nevertheless, prior work has focused on the hardware mapping of NNs rather than of modelling design variants, thus limiting the degrees of freedom. To further assist the mixed-signal architecture design process, our method extends the search space while considering hardware-related constraints and limitations. Consequently, our computer-aided design approach differs from the state-of-the-art \cite{manormachine, de_la_rosa_ai-assisted_2022, fayazi_applications_2021, nam_machine-learning_2021}. Here, relying on the IADC and RAE equivalence, the conversion process directly corresponds to a trained NN model. It offers a vast and agnostic topology exploration to identify the architecture that best maps a set of inputs to their corresponding outputs, compliant with predefined hardware-related specifications.

\subsection{Main contributions and outline}
\label{subsec:organization}

Our two main contributions are the design and optimization of complex high-order IADC architectures offered by RCNet. This $\Delta\Sigma$ IADC - RAE analogy combined with the definition of NN constraints and regularizations eases the optimizer for converter topology exploration enhancement, taking into account various hardware requirements. Section \ref{sec:RNN} presents the proposed analogy between IADC and RAE with submodel component cells, Section \ref{sec:HW} illustrates how to consider hardware-related issues for RAE model design and training, and Section \ref{sec:studyCAD} reports simulation results for DC conversion.

\section{$\Delta \Sigma$ ADC as a Recurrent AutoEncoder}
\label{sec:RNN}

Given the recursive nature of $\Delta \Sigma$ converters, the underlying idea of this work is to consider such mixed-signal structures as RAE. As depicted in Figure \ref{RAE_analogy}, the $\Delta \Sigma$ modulator can be modeled as a RNN Encoder outputting a digital sequence $\zeta$ depending on the analog input signal, while the digital filter part is similar to a RNN Decoder that enables to infer the target latent information. The proposed method relies on a supervised training based on a synthetic sample dataset. This way, the $\Delta \Sigma$ converter parameters are optimized to maximize certain performance metrics. It consists in finding optimal configurations given input-output pairs, under implementation specifications (\ie, regularizations and constraints on weights and activations) and conversion performances (\ie,  custom regression losses).

\subsection{First-order $\Delta \Sigma$ IADC: the preliminary study}
\label{subsec:SDADC}

\begin{figure}[!h]
\centering
\includegraphics[trim={0mm 45mm 0mm 0mm},clip,scale=0.45]{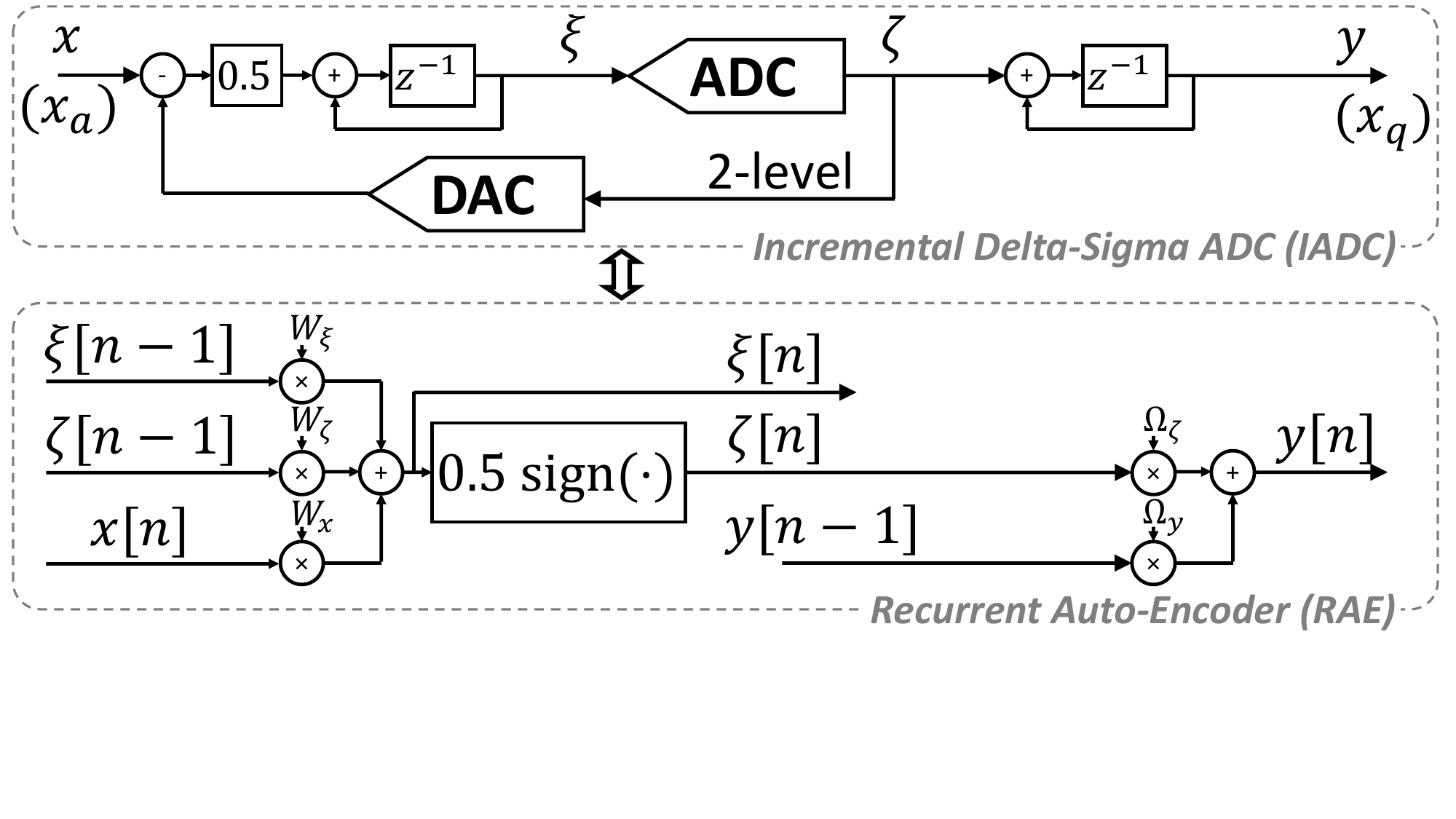}
\caption{First-order IADC and its RAE equivalence.}
\label{mod1}
\end{figure}

An analog recursive filter combined to a quantizer constitutes a $\Delta\Sigma$ modulator. It aims at shaping an oversampled continuous band-limited analog signal into a digital sequence. The key idea is the ability to reject the quantization noise out of the signal-of-interest band, allowing the use of a binary quantizer ($L$=$2$ levels, as denoted in \cite{tan_incremental_2020}). The digital filter then provides estimations $y$ of signal $x$ using the binary internal sequence $\zeta$. Figure \ref{mod1} illustrates a first-order modulator, composed of an analog integrator and a 1-bit quantizer, combined to a first-order digital integrator acting as a low-pass filter.

After the initialization (integrator and digital counter resets), a first-order $\Delta \Sigma$ IADC performs the following three operations at each conversion step, repeated $N$ times (\ie, the Over Sampling Ratio, $OSR$): $(x-\zeta)$ is integrated to provide $\xi$, then $\xi$ is quantized to update $\zeta$ and finally $\zeta$ feeds a digital filter to compute the output $y$.

In practice, this structure relies on internal signals constrained into a dynamic range centered around the quantizer threshold. This is made possible thanks to the negative feedback loop driven by the sign of the output of the integrator $\zeta$. Equations (\ref{temporal_1st_order}) and (\ref{quantiz_1st_order}) mathematically express the time domain behavior of the recursive operations, given the following four assumptions: $n\geq 1$, $|x[p]| \leq \frac { 1 } { 2 }$, $\zeta[p] = \pm \frac { 1 } { 2 } \text { for } p > 0$, and $\zeta[0] = 0.$ 

\begin{equation}
\label{temporal_1st_order}
\xi[n]=\frac { 1 } { 2 }\left(\displaystyle \sum _{p=1}^{n}x[p]-\sum _{p=0}^{n-1}\zeta[p]\right)
\end{equation}

\begin{equation} 
\label{quantiz_1st_order}
\zeta[n]=\frac { 1 } { 2 }\mathrm {s}\mathrm {i}\mathrm {g}\mathrm {n}(\xi[n])
\end{equation}

For a DC conversion of the analog level $x_a$, the quantized signal $y[N]=x_q$ outputted after $N$ conversion steps is then expressed in (\ref{Xq}) as the mean of the bitstream.

\begin{equation} 
\label{Xq}
x_{q}=\displaystyle \frac {\sum _{n=1}^{N}\zeta[n]}{N}
\end{equation}

Figure \ref{mod1} illustrates the equivalent RNN cell of the modulator, acting as described in equation (\ref{eqRNN1}) and (\ref{quantiz_1st_order}) for a single time-step. This cell takes as input $x[n]$ (current input), $\xi[n-1]$ and $\zeta[n-1]$ (hidden states), and outputs $\xi[n]$ and $\zeta[n]$.

\begin{equation} 
\label{eqRNN1}
\xi[n]=\boldsymbol{W} \boldsymbol{x}^\dagger[n] = W_{x}x[n]+W_{\zeta}\zeta[n-1]+W_{\xi}\xi[n-1]
\end{equation}

A first-order modulator is obtained choosing $W_x$=$0.5$, $W_{\zeta}$=$-0.5$ and $W_{\xi}$=$1$, where $\boldsymbol{W} = \left[W_{x}, W_{\zeta},W_{\xi}\right] \in \mathbb{R}^{1 \times 3}$ and $\boldsymbol{x}^\dagger[n]= \left[x[n]; \zeta[n-1]; \xi[n-1]\right] \in \mathbb{R}^{3 \times 1}$. On the other hand, the filter topology of the decoder is based on a SimpleRNN (SRNN) cell presented in Figure \ref{mod1}, performing the bitstream accumulation. For such a simple topology, the recurrent weight and input weight respectively are $\Omega_{\zeta}$=$1$ and $\Omega_{y}$=$1/N$, in order to provide a properly scaled output.

\subsection{High-order $\Delta \Sigma$ IADC: the RCNet generalization}
\label{subsec:GenericRNN}

\begin{figure}[!h]
\centering
\includegraphics[trim={0mm 155mm 0mm 0mm},clip,scale=0.45]{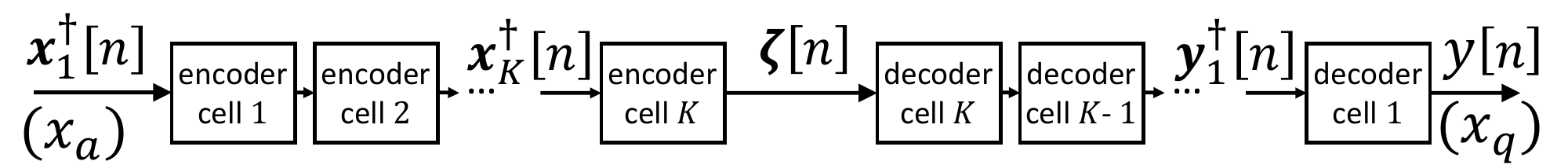}
\caption{Full model top level description using sub-cells.}
\label{full_model_cell}
\end{figure}

For a smaller quantization error (\ie, the distance between $x_a$ and $x_q$) at a given $OSR$, converters may take advantage of the features offered by high-order IADC. High-order modulators consisting in a cascade of integrators increases the order of the Noise Transfer Function (NTF), at the cost of stability issues. Using the proposed deep learning formalism, the high order structure exposed in Figure \ref{full_model_cell} may benefit from a more flexible description based on sub-cell components. 

\subsubsection{Encoding Modulator}
\label{subsec:GenericRNNMod}
 
To extend the degrees of freedom of the exploration process, we introduce a RNN cell that performs the inner product between inputs and internal weights to compute delayed and non-delayed outputs (respectively $\xi^*$ and $\xi$), as well as their delayed and non-delayed quantized counterparts (respectively $\zeta^*$ and $\zeta$). These 4 outputs typically aim to describe all the possible intra-cycle or inter-cycle types of layer connections embedded inside a modulator stage:

\setlength{\tabcolsep}{2pt}
\begin{tabular}{ l r l }
 \textbullet \hspace{1mm} Intra-cycle: & --\hspace{1mm}$\xi$: & \underline{non delayed} stage-to-stage transfer\\ 
  &  &  \underline{during} a conversion cycle \\ 
  & --\hspace{1mm}$\zeta$: & quantized $\xi$ (\textit{typ. L}=2), \\  
 \textbullet \hspace{1mm}  Inter-cycle: & --\hspace{1mm}$\xi^*$: & \underline{delayed} stage-to-stage transfer \\  
 &  & \underline{between successive} conversion cycles, \\  
 & --\hspace{1mm}$\zeta^*$: & quantized $\xi^*$ (\textit{typ. L}=2).    
\end{tabular}

Figure \ref{encoder_decoder_cell_k} details the inner product performed at each $k$ RNN cell, with the common input notation $\boldsymbol{x}^\dagger[n] \in \mathbb{R}^{4K+1}$ and its layer-specific set of weights $\boldsymbol{W}_{k}\in \mathbb{R}^{1 \times (4K+1)}$. Note that the $\textrm{cat}$ operator defines a vertical concatenation. $\zeta$ and $\xi$ variables are sequentially updated during intra-cycle operations only (not from a cycle to the next). At the beginning each cycle, $\zeta$ and $\xi$ are set to $\zeta^*$ and $\xi^*$, therefore giving for $k \in  [\![1,K]\!] $: $\zeta_k[n] = \zeta_k^*[n]$ and $\xi_k[n] = \xi_k^*[n]$. The overall modulator's weights matrix $\boldsymbol{W} \in \mathbb{R}^{K \times (4K+1)}$, for $K=3$, is reported as an example in equation (\ref{WeightGeneric}).

\begin{equation}
\label{WeightGeneric}
{\boldsymbol{W}} = 
\left[\boldsymbol{W}_{1}; \boldsymbol{W}_{2}; \boldsymbol{W}_{3}\right] =
\end{equation}
\vspace{-0.2cm}
\fontsize{10}{12}\selectfont
$$\left(
\begin{smallmatrix}
W_{01}&W_{\zeta11}&W_{\xi11}&W_{\zeta^*11}&W_{\xi^*11}&W_{\zeta21}&W_{\xi21}&W_{\zeta^*21}&W_{\xi^*21}&W_{\zeta31}&W_{\xi31}&W_{\zeta^*31}&W_{\xi^*31} \\
W_{02} & W_{\zeta12} & W_{\xi12} & W_{\zeta^*12} & W_{\xi^*12} & W_{\zeta22} & W_{\xi22} & W_{\zeta^*22} & W_{\xi^*22} & W_{\zeta32} & W_{\xi32} & W_{\zeta^*32} & W_{\xi^*32} \\
W_{03} & W_{\zeta13} & W_{\xi13} & W_{\zeta^*13} & W_{\xi^*13} & W_{\zeta23} & W_{\xi23} & W_{\zeta^*23} & W_{\xi^*23} & W_{\zeta33} & W_{\xi33} & W_{\zeta^*33} & W_{\xi^*33} \\
\end{smallmatrix}
\right)
$$
\normalsize

This generic representation offers the possibility to explore a wide range of modulator topology configurations, in which each cell has access to every cell outputs at its inputs. 

\subsubsection{Decoding Filter}
\label{subsec:GenericRNNFilt}

\begin{figure}[!h]
\centering
\includegraphics[trim={0mm 90mm 0mm 0mm},clip,scale=0.45]{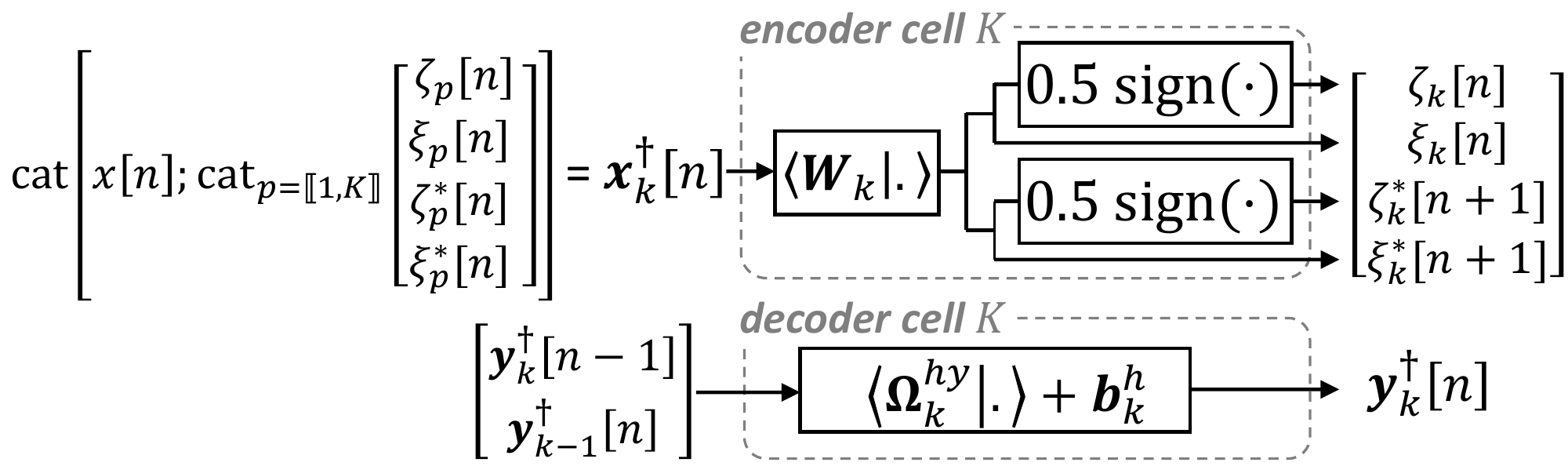}
\caption{RCNet - Encoder-Decoder cell description.}
\label{encoder_decoder_cell_k}
\end{figure}

The considered filter topology consists here in a cascade of SRNN cells, as exposed in Figure \ref{encoder_decoder_cell_k}. The first cell performs the accumulation of the incoming sequence $\zeta_K$, and each of the other cascaded cells performs the accumulation of the previous cell output. The output of each cell can be scaled to ensure a maximum normalized output for a sequence of $N$ cycles ($OSR$). In addition, in order to enable a scaled signal reconstruction at each cycle $n$, an additional normalization layer is added at the decoder output. The optimization process can thus be operated taking into account intermediate sequence lengths (\ie, $n<N$).

\section{RCNet Hardware constraints modeling}
\label{sec:HW}

To properly formulate the optimization problem, a set of hardware-related constraints and requirements has been identified, as detailed in this section and summarized in Table \ref{Tab:HW_constraints}. 

\subsection{Fidelity term optimization and stability regularization}
\label{subsec:FT}

A multi-objective loss $F$ is defined to both consider the regression problem aiming at minimizing the error between the input signal ($x_a$) and its quantized counterpart ($x_q$), so as managing stability and saturation issues. Our log-domain fidelity loss $F_{LSE}$ whose goal is to match outputs with input signals (respectively $\boldsymbol{x_q} \in \mathbb{R}^{S}$ and $\boldsymbol{x_a} \in \mathbb{R}^{S}$, with $S$ a number of training samples, equation (\ref{costFLSE})) mimics and relaxes the behavior of a max distance (\ie, a Linf norm) operator by computing the log-sum-exp of the conversion error. The additional regularization term $F_{DR}$ (equation (\ref{costDR})) adds a penalization in case of internal signals voltage excursions beyond given bounds ($\pm \delta_k$). This regularization caps the internal dynamic range to avoid saturation issues and limit signals distortion. To this end, it considers end-of-conversion inter-stage outputs $\boldsymbol{\xi} \in \mathbb{R}^{S \times K}$ corresponding to the concatenation of $K$ $\boldsymbol{\xi}_k \in \mathbb{R}^{S}$. This additional term advantageously ensures the stability of the recurrent structure at a given $OSR$ $N$, using the $\textrm {clip}_{\pm \delta_k}()$ function that trims values at input thresholds ($\pm \delta_k$), assigning outside values to boundaries. The loss function $F(\boldsymbol{x_a}, \boldsymbol{x_q}, \boldsymbol{\xi}) = F_{LSE}(\boldsymbol{x_a}, \boldsymbol{x_q}, \boldsymbol{\xi}) + \lambda_{DR} F_{DR}(\boldsymbol{\xi}), \lambda_{DR} \in \mathbb{R}$ will be computed for each batch made of $S$ samples, in which the modulator topology involves $K$ stage cells.

\begin{equation} 
\label{costFLSE}
F_{LSE}(\boldsymbol{x_a}, \boldsymbol{x_q}, \boldsymbol{\xi}) = \log \left( \log \left(\sum_{s=1}^{S} e^{\left|\boldsymbol{x_a}[s]-\boldsymbol{x_q}[s]\right|} \right)\right)
\end{equation}

\begin{equation} 
\label{costDR}
F_{DR}(\boldsymbol{\xi}) = \sum_{s=1}^{S} \sum _{k=1}^{K}\left(\boldsymbol{\xi_{k}}[s]-\textrm {clip}_{\pm \delta_k}\left (\boldsymbol{\xi_{k}}[s]\right)\right)^2
\end{equation}

\subsection{Signal saturation data activation}
\label{subsec:Noise}
In addition to $F_{DR}$ used to ensure the stability of the recurrent network while favoring a proper output scaling of the modulator stages outputs, an activation is introduced to saturate the internal signals $\xi$ and $\xi^*$ exceeding a maximum allowed amplitude. It consists in a Hardtanh layer that clips tensor values at saturation levels ($\pm \delta_k$).

\subsection{Encoder structure sparsity}
\label{subsec:Compacity}
The encoder structure sparsity is emphasized by masking latent modulator weights, simplifying the inter-layer connection map of the resulting hardware structure. A pointwise product is applied to the latent weights $\boldsymbol{W}^l$ using a learned binary mask $\boldsymbol{M} \in \left\{0,1\right\}^{K\times(4K+1)}$. $\boldsymbol{M}$ is derived from the strict heaviside of a learned latent mask $\boldsymbol{M}^l$ whose size is the same as $\boldsymbol{W}^l$. To measure this degree of sparsity, the number of Active encoder Paths $AP$ denotes the number of non-zero $\boldsymbol{W}$ weights.

\subsection{Hardware layout strategy}
\label{subsec:Quantization}
Since discrete-time $\Delta \Sigma$ modulators are commonly based on switched capacitors implementations, this paper proposes an adequate sizing that favors a low common denominator for all weight values. To this end, a Quantization Aware Training (QAT) \cite{benoit_2018} is employed to learn optimized quantized weights $\boldsymbol{W}$, being derived from latent weights $\boldsymbol{W}^l$ and $\boldsymbol{M}$. The weights are thus uniformly quantized on $Q$ absolute levels, providing $\boldsymbol{W}^i\in \mathbb{N}$. The latent weights quantization is performed during the feedforward phase, combined with a Straigh-Through Estimator for the gradient. The proposed quantization function uses the rounding to the nearest integer $\textrm{round()}$ function. The quantization step $q\in \mathbb{R}$ is a learned latent scaling, as detailed in equation (\ref{quantizfunc}). 

\begin{equation} 
\label{quantizfunc}
 \boldsymbol{W} = q \times \textrm {clip}_{\pm Q} \left(\textrm{round}\left( \frac{\boldsymbol{M} \odot \boldsymbol{W}^l }{q}\right)\right) = q \times \boldsymbol{W}^i
\end{equation}

\subsection{Thermal noise}
\label{subsec:Th Noise}
To increase the IADC robustness to temporal non-idealities, an Additional White Gaussian Noise (AWGN) is considered for in-model data augmentation activated during the training process at each internal node of the modulator. To that end, a learnable unit capacitor ${C_k}$ is assigned to each stage of the encoder. The resulting capacitors associated to each encoder stage are obtained by multiplying ${C_k}$ with respect to the $\boldsymbol{W}^i$ values. The $kTC$ noise sources due to signals sampling on capacitors can thus be simulated for each weight of the encoder. Note that this data augmentation is also activated for $SNR$ metric evaluation for validation/test results, and deactivated for $SQNR$.

\subsection{Hardware silicon surface reduction}
\label{subsec:Compacity}
An additional loss (weighted by $\lambda_{TPT}$) linearly penalizes the topologies whose the sum of capacitors ($C_{Tot}$) exceeds a Total Capacitor Threshold (TPT). This way, the learning process jointly optimizes the unit capacitors ${C_k}$ with the quantized encoder weights $\boldsymbol{W}$, promoting a balance between the silicon surface and the encoder noise robustness.

\begin{table}[!h]
\caption{Hardware-aware deep learning modeling.}
\label{Tab:HW_constraints} 
\begin{center}
\begin{tabular}{|c|c|}
\hline
\textbf{HARDWARE FEATURES} & \textbf{DEEP LEARNING MODEL}\\
\hline
\hline
\makecell{\textbf{Conversion quantization error}\\Minimize maximum conversion error} & \makecell{\textbf{Input-Output fidelity loss}\\LogSumExp (LSE) loss ($F_{LSE}$)}\\
\hline
\makecell{\textbf{Modulator Stability}\\Ensure dynamic range, limit swing} & \makecell{\textbf{Activity regularization}\\L2 on skimmed signal ($F_{DR}$)}\\
\hline
\makecell{\textbf{Operational amplifier output swing}\\ Avoid analog signal saturation} & \makecell{\textbf{Custom Activation}\\Clip with STE (\ie, Hardtanh)}\\
\hline
\makecell{\textbf{Layout implementation}\\Define a single unitary capacitor} & \makecell{\textbf{Quantization Aware Training}\\Linear weight quantization}\\
\hline
\makecell{\textbf{Thermal (kTC) Noise robustness}\\Optimize modulator noise rejection} & \makecell{\textbf{Data-Augmentation}\\Signal AWGN wrt. capacitor size}\\
\hline
\makecell{\textbf{Silicon surface and design complexity}\\Limit analytical modulator order} & \makecell{\textbf{Weight regularization}\\Favor a sparse $\boldsymbol{W}$}\\
\hline
\end{tabular}
\end{center}
\end{table}

\vspace{-0.4cm}

\section{Case study: DC conversion using IADC}
\label{sec:studyCAD}

\begin{figure}[!h]
\centering
\includegraphics[scale=1.]{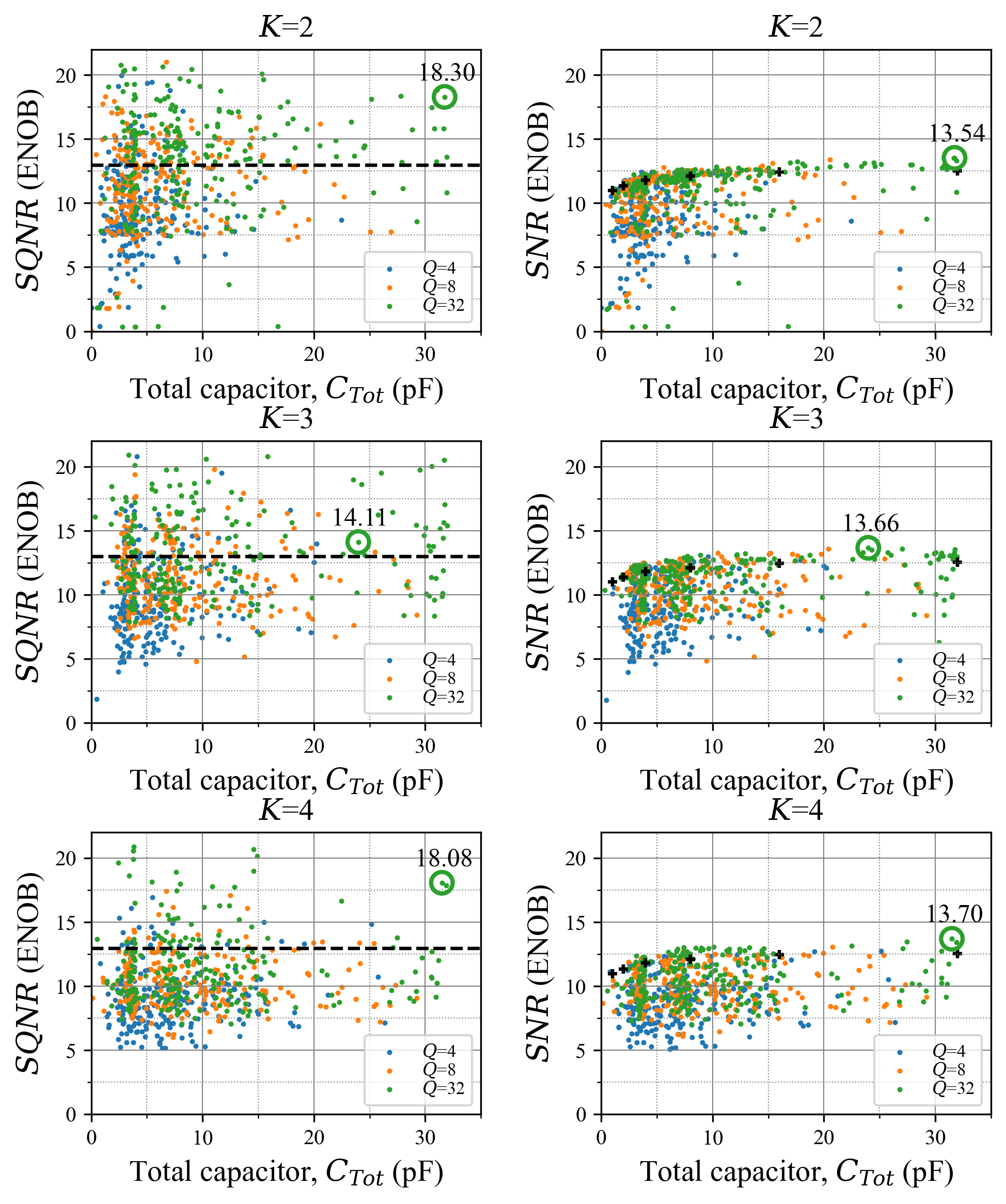}
\caption{$SQNR$ and $SNR$ at $OSR$=80 for $K\in\{2,3,4\}$ and for $Q\in\{4,8,32\}$, as a function of the total capacitor $C_{Tot}$.}
\label{SNR_totalcap}
\end{figure}

RCNet has been trained a large number of runs ($\sim$2000), under configurations listed in Table \ref{Tab:Model_parameters}, with a random initialization and decoder finetuning. To ensure a proper and stable training, the training data are generated using a uniform distribution. Figure \ref{SNR_totalcap} illustrates the resulting $SQNR$ and $SNR$, both defined as an Effective Number Of Bits (ENOB), depending on the sum of capacitors. Each training realization is reported as a colored dot depending on the weights quantization level, for different number of cells $K$. As expected, the smaller the $Q$ the worst the performance. Nevertheless, $SNR$ results for $Q$=$8$ (orange) and $Q$=$32$ (green) show that, despite weight quantization, our proposed strategy still provides better configurations compared to a CIFF IADC baseline \cite{markus_incremental_2006} (black). The best $SNR$s are encircled. Our results demonstrate a high variability depending on the model initialization. We can still observe the positive influence of the total capacitor on $SNR$, as well as the ability to find topologies with a small $C_{Tot}$.

\begin{table}[!h]
\caption{\label{param}RCNet model parameters for our simulations.}
\label{Tab:Model_parameters} 
\begin{center}
\begin{tabular}{|c|c|}
\hline
\textbf{MODEL PARAMETER TYPE} & \textbf{SET VALUES}\\
\hline
\hline
Encoder (modulator) sub-cells ($K$) &  $\left\{2, 3, 4\right\}$\\
\hline
maximum $OSR$ & 80\\
\hline
Input Dynamic Range ($x_a$)& $\in \left[-0.35,0.35\right]$\\
\hline
In-encoder saturation bounds ($\delta_k$) & 0.4\\
\hline
DAC analog levels & $\left\{-0.5, 0.5\right\}$\\
\hline
Weights quantization \#levels ($Q$) & $\left\{4, 8, 32\right\}$\\
\hline
Loss balancing ($\lambda_{DR}$, $\lambda_{TPT}$) & $0.01$, $0.0001$\\
\hline
\makecell{Total Capacitor Threshold (TPT)} & $\left\{4, 8, 16, 32\right\}$\\
\hline
\makecell{Temporal sampling noise} & Cell-wise optimized kT/C \\
\hline
\end{tabular}
\end{center}
\end{table}

\begin{figure}[!h]
\centering
\includegraphics[scale=1.]{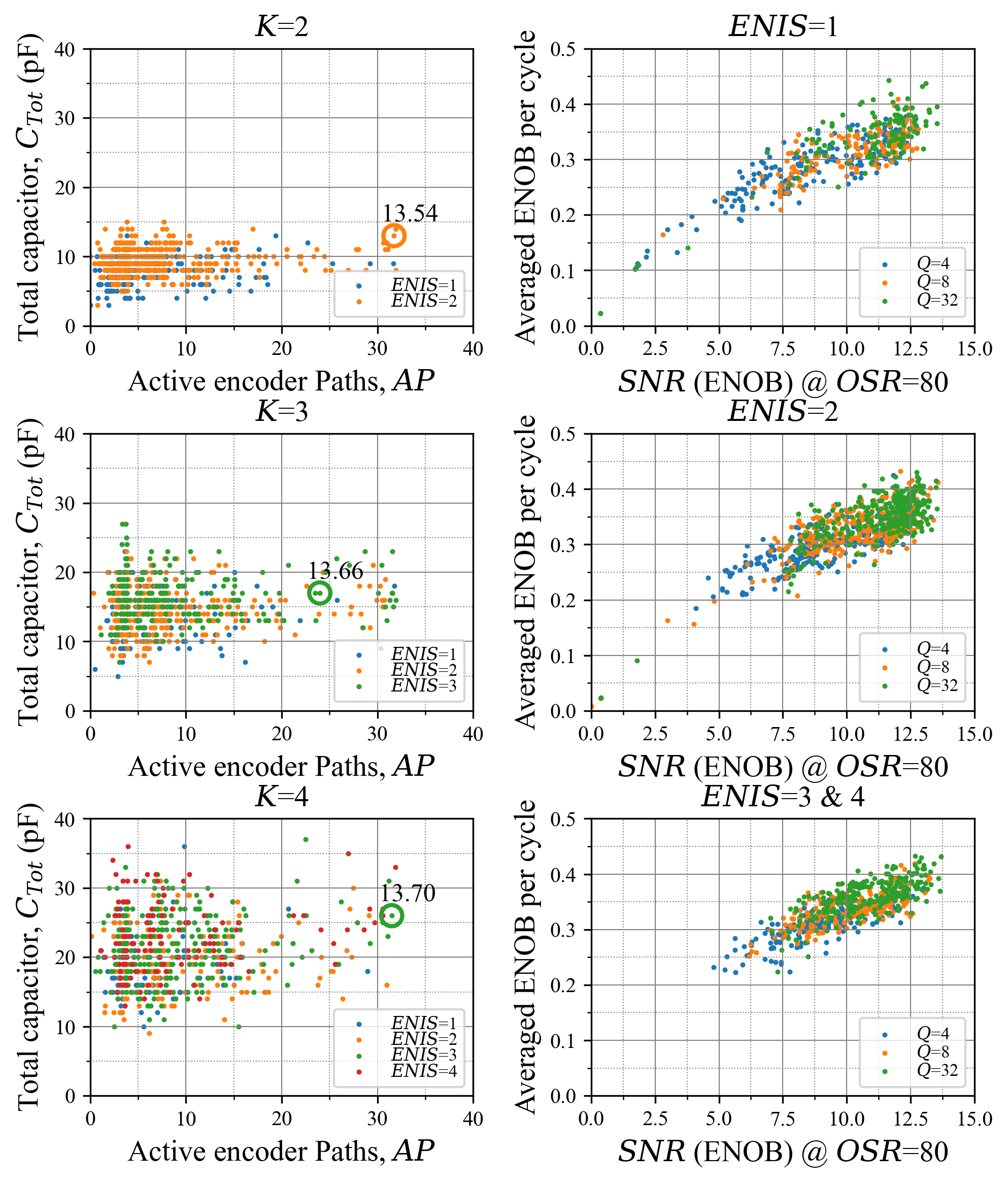}
\caption{Total capacitor $C_{Tot}$ versus the number of Active encoder Paths $AP$, and average ENOB per cycle versus simulated $SNR$ for $OSR$ from 1 to 80.}
\label{SNR_bitpercycle}
\end{figure}

\begin{figure}[!h]
\centering
\includegraphics[scale=1.]{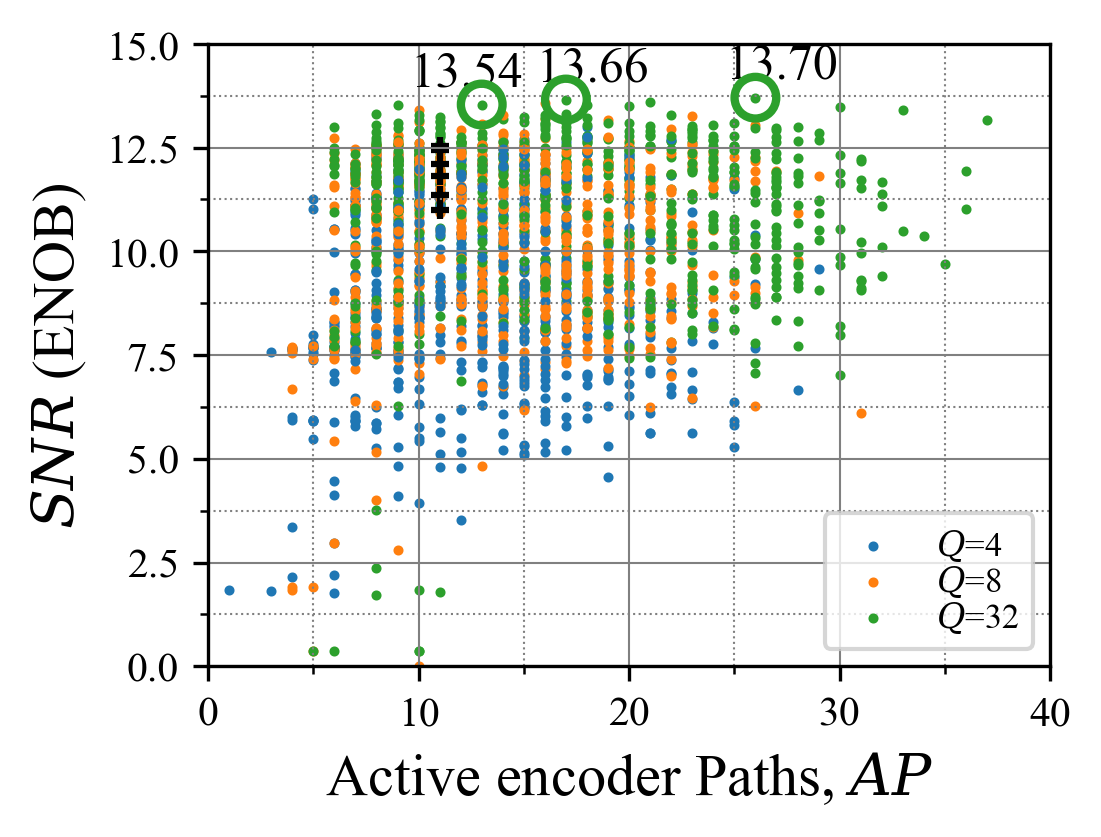}
\caption{$SNR$ for different encoder orders $K$ and weight quantization levels $Q$, as a function of the number of Active encoder Paths ($AP$).}
\label{SNR_activepath}
\end{figure}

In Figure \ref{SNR_bitpercycle}, the Effective Number of Integration Stages ($ENIS$) characterizes the hardware complexity associated to each learned RCNet. On the other hand, Figure \ref{SNR_activepath} reports the $SNR$ as a function of $AP$. A small $Q$ limits the ability to explore complex configurations. To compute $ENIS$, a stage $k$ defines an integration stage if one of the weights $W_{\xi kk}$ or $W_{\xi^*kk}$ is different from zero. Figure \ref{SNR_bitpercycle} demonstrates that the best $SNR$s are not necessarily linked to the highest total capacitor or number of active paths (even $ENIS=K$), demonstrating the ability of our learning strategy to balance design constraints versus the $SNR$. The average ENOB per cycle with respect to the $SNR$ at $OSR$=$80$ (right column, Figure \ref{SNR_bitpercycle}) attempts to characterize each topology depending on its convergence trend, highlighting the conversion speed. While a low $ENIS$ corresponds to a low-order architecture, which requires a larger $OSR$ to reach a given $SNR$, it appears here that these kind of topologies can even compete with higher-order. It denotes the ability of this strategy to provide unconventional signal shaping such as exponential-like decay integration scheme \cite{patent_DWI, paper_DWI, spoiler}, which enables the reduction of quantization noise for a given $OSR$.

\section{Conclusion}
\label{sec:conclusion}
Based on the analogy between an IADC and a RAE, our proposed RCNet deep learning methodology enables an hardware-aware topology exploration, jointly optimizing different design specifications. The proposed method does not rely on any \textit{a priori} assumptions and can provide architectures offering unique tradeoffs between $SNR$ and hardware constraints. It has also been found that this strategy converges towards topologies that take advantage of unconventional signal shaping. It thus constitutes a powerful tool to investigate innovative topologies difficult to size using standard analytical approaches. Future works will extend to converters aiming at directly estimating latent variables (such as signal frequency), constituting a step towards Analog-to-Information converters.

\bibliographystyle{IEEEtran}
\bibliography{AI-designedADC}

\begin{thebibliography}{10}
\providecommand{\url}[1]{#1}
\csname url@samestyle\endcsname
\providecommand{\newblock}{\relax}
\providecommand{\bibinfo}[2]{#2}
\providecommand{\BIBentrySTDinterwordspacing}{\spaceskip=0pt\relax}
\providecommand{\BIBentryALTinterwordstretchfactor}{4}
\providecommand{\BIBentryALTinterwordspacing}{\spaceskip=\fontdimen2\font plus
\BIBentryALTinterwordstretchfactor\fontdimen3\font minus
  \fontdimen4\font\relax}
\providecommand{\BIBforeignlanguage}[2]{{%
\expandafter\ifx\csname l@#1\endcsname\relax
\typeout{** WARNING: IEEEtran.bst: No hyphenation pattern has been}%
\typeout{** loaded for the language `#1'. Using the pattern for}%
\typeout{** the default language instead.}%
\else
\language=\csname l@#1\endcsname
\fi
#2}}
\providecommand{\BIBdecl}{\relax}
\BIBdecl

\bibitem{markus_incremental_2006}
J.~Markus \emph{et~al.}, ``Incremental {Delta}-{Sigma} {Structures} for {DC}
  {Measurement}: an {Overview},'' in \emph{{IEEE} {Custom} {Integrated}
  {Circuits} {Conference} 2006}, Sep. 2006, pp. 41--48.

\bibitem{tan_incremental_2020}
Z.~Tan \emph{et~al.}, ``Incremental {Delta}-{Sigma} {ADCs}: {A} {Tutorial}
  {Review},'' \emph{IEEE Transactions on Circuits and Systems I: Regular
  Papers}, vol.~67, no.~12, pp. 4161--4173, Dec. 2020.

\bibitem{markus_theory_2004}
J.~Markus \emph{et~al.}, ``Theory and applications of incremental
  {$\Delta$}{$\Sigma$} converters,'' \emph{IEEE Transactions on Circuits and
  Systems I: Regular Papers}, vol.~51, no.~4, pp. 678--690, Apr. 2004.

\bibitem{de_la_rosa_sigma-delta_2011}
J.~M. de~la Rosa, ``Sigma-{Delta} {Modulators}: {Tutorial} {Overview}, {Design}
  {Guide}, and {State}-of-the-{Art} {Survey},'' \emph{IEEE Transactions on
  Circuits and Systems I: Regular Papers}, vol.~58, no.~1, pp. 1--21, Jan.
  2011.

\bibitem{kieu_outlier_2019}
T.~Kieu \emph{et~al.}, ``Outlier {Detection} for {Time} {Series} with
  {Recurrent} {Autoencoder} {Ensembles},'' in \emph{International Joint
  Conference on Artificial Intelligence}, 2019, pp. 2725--2732.

\bibitem{xie_anomaly_2023}
T.~Xie \emph{et~al.}, ``Anomaly detection for multivariate times series through
  the multi-scale convolutional recurrent variational autoencoder,''
  \emph{Expert Systems with Applications}, vol. 231, p. 120725, Nov. 2023.

\bibitem{weninger_deep_2014}
F.~Weninger \emph{et~al.}, ``Deep recurrent de-noising auto-encoder and blind
  de-reverberation for reverberated speech recognition,'' in \emph{{IEEE}
  {International} {Conference} on {Acoustics}, {Speech} and {Signal}
  {Processing}}, May 2014, pp. 4623--4627.

\bibitem{hou_lstm-based_2020}
B.~Hou \emph{et~al.}, ``{LSTM}-{Based} {Auto}-{Encoder} {Model} for {ECG}
  {Arrhythmias} {Classification},'' \emph{IEEE Transactions on Instrumentation
  and Measurement}, vol.~69, no.~4, pp. 1232--1240, Apr. 2020.

\bibitem{yang_learning_2021}
R.~Yang \emph{et~al.}, ``Learning for {Video} {Compression} {With} {Recurrent}
  {Auto}-{Encoder} and {Recurrent} {Probability} {Model},'' \emph{IEEE Journal
  of Selected Topics in Signal Processing}, vol.~15, no.~2, pp. 388--401, Feb.
  2021.

\bibitem{wang_comprehensive_2022}
Q.~Wang \emph{et~al.}, ``\BIBforeignlanguage{en}{A {Comprehensive} {Survey} of
  {Loss} {Functions} in {Machine} {Learning}},''
  \emph{\BIBforeignlanguage{en}{Annals of Data Science}}, vol.~9, no.~2, pp.
  187--212, Apr. 2022.

\bibitem{dubey_activation_2022}
S.~R. Dubey \emph{et~al.}, ``Activation functions in deep learning: {A}
  comprehensive survey and benchmark,'' \emph{Neurocomputing}, vol. 503, pp.
  92--108, Sep. 2022.

\bibitem{thiruvathukal_survey_2022}
G.~K. Thiruvathukal \emph{et~al.}, ``\BIBforeignlanguage{en}{A {Survey} of
  {Quantization} {Methods} for {Efficient} {Neural} {Network} {Inference}},''
  pp. 291--326, Jan. 2022, book Title: Low-Power Computer Vision Edition: 1
  ISBN: 9781003162810 Place: Boca Raton Publisher: Chapman and Hall/CRC.

\bibitem{mumuni_data_2022}
A.~Mumuni \emph{et~al.}, ``Data augmentation: {A} comprehensive survey of
  modern approaches,'' \emph{Array}, vol.~16, p. 100258, Dec. 2022.

\bibitem{Xu2018AnalogtodigitalCR}
S.~Xu \emph{et~al.}, ``Analog-to-digital conversion revolutionized by deep
  learning,'' \emph{arXiv: Signal Processing}, 2018.

\bibitem{samiee_deep_2021}
A.~Samiee \emph{et~al.}, ``\BIBforeignlanguage{en}{Deep analog-to-digital
  converter for wireless communication},'' in
  \emph{\BIBforeignlanguage{en}{{AI} and {Optical} {Data} {Sciences} {II}}},
  K.-i. Kitayama \emph{et~al.}, Eds.\hskip 1em plus 0.5em minus 0.4em\relax
  Online Only, United States: SPIE, Mar. 2021, p.~44.

\bibitem{bansal_neural-network_2021}
S.~Bansal \emph{et~al.}, ``Neural-{Network} {Based} {Self}-{Initializing}
  {Algorithm} for {Multi}-{Parameter} {Optimization} of {High}-{Speed}
  {ADCs},'' \emph{IEEE Transactions on Circuits and Systems II: Express
  Briefs}, vol.~68, no.~1, pp. 106--110, Jan. 2021.

\bibitem{de_la_rosa_ai-assisted_2022}
J.~M. de~la Rosa, ``{AI}-{Assisted} {Sigma}-{Delta}
  {Converters}—{Application} to {Cognitive} {Radio},'' \emph{IEEE
  Transactions on Circuits and Systems II: Express Briefs}, vol.~69, no.~6, pp.
  2557--2563, Jun. 2022.

\bibitem{fayazi_applications_2021}
M.~Fayazi \emph{et~al.}, ``Applications of {Artificial} {Intelligence} on the
  {Modeling} and {Optimization} for {Analog} and {Mixed}-{Signal} {Circuits}:
  {A} {Review},'' \emph{IEEE Transactions on Circuits and Systems I: Regular
  Papers}, vol.~68, no.~6, pp. 2418--2431, Jun. 2021.

\bibitem{nam_machine-learning_2021}
J.-W. Nam \emph{et~al.}, ``Machine-{Learning} based {Analog} and {Mixed}-signal
  {Circuit} {Design} and {Optimization},'' in \emph{{International}
  {Conference} on {Information} {Networking}}, Jan. 2021, pp. 874--876.

\bibitem{tank_1986}
D.~Tank \emph{et~al.}, ``Simple 'neural' optimization networks: {An A/D}
  converter, signal decision circuit, and a linear programming circuit,''
  \emph{IEEE Transactions on Circuits and Systems}, vol.~33, no.~5, pp.
  533--541, 1986.

\bibitem{cao_neuadc_2020}
W.~Cao \emph{et~al.}, ``{NeuADC}: {Neural} {Network}-{Inspired} {Synthesizable}
  {Analog}-to-{Digital} {Conversion},'' \emph{IEEE Transactions on
  Computer-Aided Design of Integrated Circuits and Systems}, vol.~39, no.~9,
  pp. 1841--1854, Sep. 2020.

\bibitem{james_neural_2018}
A.~Tankimanova \emph{et~al.}, ``\BIBforeignlanguage{en}{Neural
  {Network}-{Based} {Analog}-to-{Digital} {Converters}},'' in
  \emph{\BIBforeignlanguage{en}{Memristor and {Memristive} {Neural}
  {Networks}}}, A.~P. James, Ed.\hskip 1em plus 0.5em minus 0.4em\relax InTech,
  Apr. 2018.

\bibitem{danial_pipelined_2020}
L.~Danial \emph{et~al.}, ``A {Pipelined} {Memristive} {Neural} {Network}
  {Analog}-to-{Digital} {Converter},'' in \emph{{IEEE} {International}
  {Symposium} on {Circuits} and {Systems}}, Oct. 2020, pp. 1--5.

\bibitem{danial_delta-sigma_2019}
------, ``Delta-{Sigma} {Modulation} {Neurons} for {High}-{Precision}
  {Training} of {Memristive} {Synapses} in {Deep} {Neural} {Networks},'' in
  \emph{{IEEE} {International} {Symposium} on {Circuits} and {Systems}}, May
  2019, pp. 1--5.

\bibitem{manormachine}
J.~Wagner \emph{et~al.}, ``Man or machine — design automation of delta-sigma
  modulators,'' in \emph{2018 IEEE International Symposium on Circuits and
  Systems (ISCAS)}, 2018, pp. 1--5.

\bibitem{benoit_2018}
B.~Jacob \emph{et~al.}, ``{Quantization and Training of Neural Networks for
  Efficient Integer-Arithmetic-Only Inference},'' in \emph{IEEE/CVF Conference
  on Computer Vision and Pattern Recognition}, 2018, pp. 2704--2713.

\bibitem{patent_DWI}
A.~Verdant \emph{et~al.}, ``High-linearity sigma-delta converter,''
  \emph{Patent WOEP2016/053687}, 2016.

\bibitem{paper_DWI}
W.~Guicquero \emph{et~al.}, ``{Incremental Delta Sigma Modulation with Dynamic
  Weighted Integration},'' in \emph{IEEE International Midwest Symposium on
  Circuits and Systems}, 2018, pp. 344--347.

\bibitem{spoiler}
B.~Wang \emph{et~al.}, ``{A 550 µW 20 kHz BW 100.8 DB SNDR linear-exponential
  multi-bit incremental converter with 256-cycles in 65 nm CMOS},'' in
  \emph{IEEE Symp. VLSI Circuits}, 2018, pp. 207--208.

\end{thebibliography}

\end{document}